\documentstyle[seceq]{ptptex}

\newcommand{\wt}{\widetilde}
\newcommand{\wh}{\widehat}

\newcommand{\U}{{\rm U}}
\newcommand{\SO}{{\rm SO}}
\newcommand{\SU}{{\rm SU}}

\newcommand{\rap}[2]
{\setbox1=\hbox{#1}%
\setbox2=\hbox to\wd1{\hss #2\hss}%
\mbox{\rlap{\box1}\box2}}
\newcommand{\sla}[1]{\rap{$#1$}{/}}

\title{
1/4 BPS solutions in massive IIA supergravity
}

\author{
Yosuke {\sc Imamura}\footnote{E-mail: imamura@hep-th.phys.s.u-tokyo.ac.jp}
}

\inst{
Department of Physics, University of Tokyo, Tokyo 113-0033, Japan}

\abst{
We study four kinds of $1/4$ BPS solutions in massive IIA supergravity
corresponding to
D8-D0-F1, D8-D2, D8-D4 and D8-D6-NS5 systems.
We show that these solutions
are reproduced without making nontrivial assumptions
by using supersymmetry conditions.
D8-D2 and D8-D4 solutions are represented by harmonic functions, as usual,
while the other two are represented by solutions
of non-linear differential equations.
Because these four solutions can be treated in almost identical ways,
we mainly focus on the D8-D6-NS5 systems.
We first discuss D6-NS5 solutions with uniform mass parameters.
Then, we introduce D8-branes as domain walls by connecting two
solutions with different values of the mass parameter.
We also discuss boundary conditions and supersymmetry on domain walls.
}

\begin{document}

\maketitle

\section{Introduction}
Dualities between string theories and gauge theories are important subjects
in the recent development of string theories.
Because gauge fields reside on branes,
we can investigate dynamics of gauge theories by studying
brane configurations in a certain limit in which decouples gravity is decoupled.
In this framework, classical solutions in supergravities play important roles.
By analyzing the geometry of the solutions,
we can study non-perturbative dynamics of gauge theories
realized on branes in the background.

In this paper, we study four kinds of 
1/4 BPS solutions including D8-branes:
D8-D0-F1, D8-D2, D8-D4 and D8-D6-NS5 solutions.
Because we can treat them in almost the same way,
we mainly focus on D8-D6-NS5 systems.
In the context of supergravity, D8-branes are regarded as domain walls separating
regions with different values of the mass parameter $m$,
whose square is the cosmological constant.
Therefore, when we treat D8-branes in supergravity,
we should necessarily study classical solutions in the massive IIA supergravity
proposed by Romans\cite{massiveiia}.

Ordinary BPS solutions consisting of $n$ kinds of branes
are represented by $n$ harmonic functions.
As specific such examples, D8-D2 and D8-D4 solutions are represented by
two harmonic functions.\cite{bbj,gkt,youm}
However,
as is mentioned in Ref.\citen{janssen},
D8-D0-F1 and D8-D6-NS5 solutions
have different structure.
A reason for this is as follows.
One important feature of massive IIA supergravity
is the correlation between charges of different kinds of branes.
In massive backgrounds,
D0-branes and NS5-branes cannot exist by themselves.
They are always accompanied by fundamental strings and D6-branes,
respectively.
Let us discuss the NS5 and the D6-brane charges.
These charges are defined as sources of the NS-NS $3$-form field
strength $H_3=dB_2$
and the R-R $2$-form field strength $G_2=dC_1+mB_2$, respectively.
The gauge transformation of $B_2$ also transforms the $C_1$ field:
\begin{equation}
\delta B_2=d\Lambda_1,\quad
\delta C_1=-m\Lambda_1.
\end{equation}
By this gauge transformation, we can always gauge away $C_1$,
and the action given in Ref.\citen{massiveiia} does not contain the field $C_1$.
We, however, do not eliminate $C_1$ in order
to allow investigation of the $m=0$ case simultaneously.
The Bianchi identity for the field strength $G_2$ is
\begin{equation}
dG_2=mH_3.
\label{g2h3}
\end{equation}
If we introduce the D6-brane charge density as $\rho_{\rm D6}=dG_2$
and the NS5-brane charge density as $\rho_{\rm NS5}=dH_3$,
we see from (\ref{g2h3}) that these are related by
\begin{equation}
d\rho_{\rm D6}=m\rho_{\rm NS5}.
\label{twocharges}
\end{equation}
Mathematically, these charges are classified as gerbes\cite{gerbes}.
Due to this correlation of the charges,
a massive D6-NS5 solution is represented by
only one function determined by the D6-brane charge density $\rho_{\rm D6}$.
We will explicitly show that this is indeed the case.

The relation (\ref{twocharges})
is closely related to the Hanany-Witten effect\cite{HW}.
If an NS5-brane moves from a region with a mass parameter $m=m_0$
to a region with $m=m_0+1$,
the number of D6-branes ending on the NS5-brane changes by $1$,
and a new D6-brane stretched between the NS5-brane
and the D8-brane must be generated.

Almost all of these solutions are already known.
A large class of intersecting brane solutions,
including D8-D2 and D8-D4 solutions,
are constructed in Refs.\cite{bbj} and \cite{youm}.
The D8-D0-F1 solution presented in this paper is essentially equivalent
to what is obtained in Ref.\citen{081}.
D8-D0-F1 and D8-D6-NS5 solutions are also constructed in Ref.\citen{janssen}
for a special case.

This paper is organized as follows.
In Section \ref{568.sec}, we present a D6-NS5 solution of massive IIA supergravity
with a uniform mass parameter.
A derivation of the solution is given in Section \ref{massived6.sec}.
There, we find a $1/4$ BPS classical solution
by means of supersymmetry conditions, which guarantee the existence of
unbroken rigid supersymmetries.
In section \ref{domainwall.sec}, we discuss
the construction of D8-branes as domain walls
by connecting two solutions with different values of the mass parameter.
We also discuss the BPS condition on domain walls.
Several concrete examples are given in Section \ref{examples.sec}.
We briefly discuss a D8-D0-F1 solution in Section \ref{018.sec}
and D8-D2 and D8-D4 solutions in Section \ref{2848.sec}.
The last section is devoted to discussions.
Our conventions and notation are summarized in Appendix A.
\section{D6-NS5 solution}\label{568.sec}
In this section, we give a supergravity solution representing a system
with NS5-branes along the $x^0,\ldots,x^5$ directions
and D6-branes along the $x^0,\ldots,x^5,z$ directions.
This system preserves $1/4$ supersymmetry specified by
\begin{equation}
\Gamma^{\wh z}\epsilon=\epsilon,\quad
\Gamma^{\wh r\wh\theta^1\wh\theta^2}\Gamma_{11}\epsilon=\epsilon.
\label{susycond}
\end{equation}
In addition to these two kinds of branes,
we can add D8-branes perpendicular to the $z$ direction
while maintaining the $1/4$ supersymmetry
(Table \ref{568.tbl}).
From the viewpoint of supergravity, D8-branes are regarded as domain walls separating
regions with different values of the mass parameter.
We first discuss solutions with uniform mass parameters
and later consider connecting different solutions
to make domain wall solutions.
\begin{table}[htb]
\caption{$1/4$ BPS configuration consisting of D8, D6 and NS5-branes.}
\label{568.tbl}
\begin{center}
\begin{tabular}{ccccccccccc}
\hline
\hline
           & $x^0$ & $x^1$ & $x^2$ & $x^3$ & $x^4$ & $x^5$ & $z$ & $r$ & $\theta^1$ & $\theta^2$ \\
\hline
\mbox{D8}  & $\circ$ & $\circ$ & $\circ$ & $\circ$ & $\circ$ & $\circ$ && $\circ$ & $\circ$ & $\circ$ \\
\mbox{D6}  & $\circ$ & $\circ$ & $\circ$ & $\circ$ & $\circ$ & $\circ$ & $\circ$\\
\mbox{NS5} & $\circ$ & $\circ$ & $\circ$ & $\circ$ & $\circ$ & $\circ$ \\
\hline
\end{tabular}
\end{center}
\end{table}
We assume Poincar\'e symmetry along the $x^i$ ($i=0,\ldots,5$) directions,
rotational symmetry and parity symmetry in the $(r,\theta^1,\theta^2)$ plane,
and the $1/4$ supersymmetry given by (\ref{susycond}).
As we show in the next section,
any solution
possessing these symmetries can be written in the form
\begin{eqnarray}
&&ds^2=S^{-1/2}\eta_{ij}dx^idx^j+KS^{-1/2}dz^2
            +KS^{1/2}(dr^2+r^2d\Omega_2^2),
\label{568metric}\\
&&e^\phi=g_{\rm str}K^{1/2}S^{-3/4},\label{568dilaton}\\
&&G_{\theta^1\theta^2}=-\frac{r^2}{g_{\rm str}}\sqrt{\gamma}
           \frac{\partial S}{\partial r},
\label{rr2}\\
&&H_{z\theta^1\theta^2}=-r^2\sqrt{\gamma}\frac{\partial K}{\partial r},\quad
H_{r\theta^1\theta^2}=r^2\sqrt{\gamma}\frac{\partial}{\partial z}(KS),
\label{NSNS3}
\end{eqnarray}
where $S$ and $K$ are functions of $z$ and $r$.
We write the metric on the unit ${\bf S}^2$ as
\begin{equation}
d\Omega_2^2=\gamma_{\theta^m\theta^n}d\theta^md\theta^n.\quad(m,n=1,2)
\end{equation}
The quantity $\gamma$ in (\ref{rr2}) and (\ref{NSNS3})
is the determinant of $\gamma_{\theta^m\theta^n}$.
(We use the symbol $\theta^m$ to represent
both an index for angular coordinates and the
angular coordinates themselves.)
If we introduce the dual fields $C_7$ and $B_6$ by
\begin{equation}
G_2=-*dC_7,\quad
H_3=-e^{2\phi}*(dB_6-mC_7),
\end{equation}
Eqs. (\ref{rr2}) and (\ref{NSNS3}) can be replaced by simpler ones:
\begin{equation}
C_{012345z}=\frac{1}{g_{\rm str}S},\quad
B_{012345}=\frac{1}{g_{\rm str}^2K}.
\label{C7B6}
\end{equation}
The function $S$ is a solution of
\begin{equation}
\Delta_3S+\frac{1}{2}\frac{\partial^2}{\partial z^2}S^2=0,
\label{Sequation}
\end{equation}
where $\Delta_3$ is the Laplacian in the $(r,\theta^1,\theta^2)$ plane
defined by
\begin{equation}
\Delta_3=\frac{1}{r^2}\frac{\partial}{\partial r}r^2\frac{\partial}{\partial r}.
\label{laplacian}
\end{equation}
The $\theta^m$-derivative terms are absent because we consider only
rotationally symmetric solutions.

The functions $S$ and $K$ satisfy the relation
\begin{equation}
mg_{\rm str}K=\frac{\partial S}{\partial z}.
\label{KSeq}
\end{equation}
When $m\neq0$, we can determine the function $K$ by this equation.
This corresponds to the fact that the NS5-brane charge density
in massive backgrounds is uniquely determined by the D6-brane charge density.

If $m=0$, Eq. (\ref{KSeq}) implies that
$S$ is independent of the coordinate $z$,
and the differential equation (\ref{Sequation}) is solved by
\begin{equation}
S=1+\frac{Q_6g_{\rm str}}{4\pi r}.
\label{masslessS}
\end{equation}
Because we can put NS5-branes in massless backgrounds
without changing the D6-brane charge density,
we need to determine the function $K$ separately
by solving the Laplace equation
\begin{equation}
\Delta_3K+S\frac{\partial^2}{\partial z^2}K=0.
\label{Kequation}
\end{equation}
This equation is formally obtained from (\ref{Sequation})
by differentiating it with respect to $z$ and taking the limit $m\rightarrow0$.

\section{Derivation of the solution}\label{massived6.sec}
In this section, we derive the D6-NS5 solution given in the previous section.
We assume that the solution possesses
Poincare symmetry on $(x^0,\ldots,x^5)$,
rotational and parity symmetry on $(r,\theta^1,\theta^2)$,
and $1/4$ supersymmetry (\ref{susycond}).
We can take the metric in the form
\begin{equation}
ds^2
=f^2(r,z)\eta_{ij}dx^idx^j+g^2(r,z)dz^2
+h^2(r,z)dr^2+\rho^2(r,z)d\Omega_2^2.
\label{metric}
\end{equation}
On any two-dimensional surface a locally orthogonal coordinate system
can be defined,
and we have adopted such a coordinate system on the $z$-$r$ plane.
Non-zero components of the spin connection
in the spacetime with the metric (\ref{metric}) are
\begin{eqnarray}
&&
\omega_{\wh i-\wh j\wh z}=\eta_{\wh i\wh j}\frac{f_{\wh z}}{f},\quad
\omega_{\wh\theta^m-\wh\theta^n\wh z}
     =\delta_{\wh\theta^m\wh\theta^n}\frac{\rho_{\wh z}}{\rho},\quad
\omega_{\wh r-\wh r\wh z}=\frac{h_{\wh z}}{h},\nonumber\\
&&
\omega_{\wh i-\wh j\wh r}=\eta_{\wh i\wh j}\frac{f_{\wh r}}{f},\quad
\omega_{\wh\theta^m-\wh\theta^n\wh r}
     =\delta_{\wh\theta^m\wh\theta^n}\frac{\rho_{\wh r}}{\rho},\quad
\omega_{\wh z-\wh z\wh r}=\frac{g_{\wh r}}{g},\nonumber\\
&&\omega_{\wh\theta^k-\wh\theta^m\wh\theta^n}
=\frac{1}{\rho}\wt\omega_{\wh\theta^k-\wh\theta^m\wh\theta^n},
\end{eqnarray}
where hatted indices are local Lorentz indices.
Here $\wt\omega_{\wh\theta^k-\wh\theta^m\wh\theta^n}$ is the
spin connection on the unit ${\bf S}^2$ made from $\gamma_{mn}$.
A local Lorentz frame is chosen so that the vielbein becomes diagonal.
We use subscripts to represent derivatives
(i.e. $f_{\wh z}\equiv\partial_{\wh z}f=e^z_{\wh z}\partial_zf$).

By the symmetries we have assumed,
the only non-zero components of the $2$-form fields are
\begin{equation}
B_{\theta^1\theta^2}=\sqrt{\gamma}b(z,r),\quad
G_{\theta^1\theta^2}=\sqrt{\gamma}c(z,r).
\label{bc}
\end{equation}
$B_{zr}$ and $G_{zr}$ are forbidden by parity symmetry
in the $(r,\theta^1,\theta^2)$ plane.
If $m=0$, the function $c$ is a constant, due to the flux conservation law.
If $m\neq0$, we can eliminate the R-R $1$-form field
$C_1$ by a $B$-field gauge transformation,
and the two functions in (\ref{bc}) are related by $c=mb$.
We do not use these facts at this point,
in order to allow investigation of the massless case
and the massive case simultaneously.
Although the potential $B_{mn}$ does not have to be invariant under the $\SO(3)$ rotation along
$\theta^m$, we can use a gauge with an $\SO(3)$ invariant potential.

Substituting (\ref{metric}) and (\ref{bc}) into the equations of motion
for massive IIA supergravity\cite{massiveiia}
and solving them,
we would obtain classical solutions.
However, the equations of motion are second-order differential equations,
and they are not easy to solve.
Because we want BPS solutions,
we can use the supersymmetry conditions
$\delta\psi_m=\delta\lambda=0$ instead of the equations of motion.
They are first-order differential equations,
and it is quite easy for us to solve them.
Therefore, we use the supersymmetry conditions to obtain the classical solutions.
The supersymmetry transformation laws of the fermion fields are\cite{massiveiia}
\begin{eqnarray}
\delta\psi_m
&=&D_m\epsilon
-\frac{1}{4\cdot2}H_{mnp}\Gamma^{np}\Gamma_{11}\epsilon
+\frac{e^\phi}{8}\sla{G}\Gamma_m\epsilon,\\
\delta\lambda
&=&\Gamma^mD_m\epsilon
-(\partial_m\phi)\Gamma^m\epsilon
-\frac{1}{4}\sla{H}_3\Gamma_{11}\epsilon,
\end{eqnarray}
up to $3$-fermion terms.
(For our conventions and notation, see Appendix \ref{convention.sec}.)

Let us fix the $z$ direction and the $r$ direction in the $z$-$r$ plane
using the supersymmetry parameter $\epsilon$.
We have assumed that the solution is 1/4 BPS.
In the limit in which all brane charges vanish,
$\epsilon$ becomes a constant spinor
satisfying (\ref{susycond}) in the flat spacetime.
Thus, $\epsilon$ for a curved solution must satisfy the conditions
\begin{equation}
v_{\wh\alpha}\Gamma^{\wh\alpha}\epsilon=\epsilon,\quad
w_{\wh\alpha}\Gamma^{\wh\alpha\wh\theta^1\wh\theta^2}
                          \Gamma^{11}\epsilon=\epsilon,
\end{equation}
where $v$ and $w$ are vector fields in the $z$-$r$ plane
satisfying the ortho-normal conditions $v^2=w^2=1$ and $v\cdot w=0$.
We choose the directions of these two vectors as the $z$ axis and the $r$ axis.
Then, the parameter $\epsilon$ satisfies (\ref{susycond})
at any point in the spacetime.

The parameter $\epsilon$ of supersymmetry transformations
can be decomposed as
\begin{equation}
\epsilon=s(z,r)\epsilon_0(\theta^m),
\end{equation}
where $s$ and $\epsilon_0$ are a scalar and a spinor function, respectively.
Due to rotational symmetry, the angular part $\epsilon_0$ must coincide
with that of a constant spinor
on a flat spacetime, which satisfies
\begin{equation}
0=D^{\rm flat}_{\wh\theta^m}\epsilon_0
={\cal D}_{\wh\theta^m}\epsilon_0+\frac{1}{2\rho}\Gamma^{\wh\theta^m\wh r}\epsilon_0,
\label{byflat}
\end{equation}
where ${\cal D}_{\wh\theta^m}$
is the covariant derivative on ${\bf S}^2$ with radius $\rho$,
which does not include the contribution
of the component $\omega_{\theta^i-\theta^jr}$ of the spin connection.
We can use Eq. (\ref{byflat}) to eliminate $\theta^m$-derivatives of
$\epsilon$ in the supersymmetry conditions.

Substituting these expressions into the supersymmetry conditions,
we obtain five independent equations:
\begin{eqnarray}
\Gamma_{\wh1}\delta\psi_{\wh1}
&=&
 \left(\frac{1}{2}\frac{f_{\wh z}}{f}+\frac{e^\phi}{8}m\right)\epsilon
+\left(\frac{1}{2}\frac{f_{\wh r}}{f}-\frac{e^\phi}{8}\frac{c}{\rho^2}\right)\Gamma^{\wh r}\epsilon=0,\\
\Gamma_{\wh z}\delta\psi_{\wh z}
&=&\left(\frac{s_{\wh z}}{s}+\frac{e^\phi}{8}m\right)\epsilon
  +\left(\frac{1}{2}\frac{g_{\wh r}}{g}
  +\frac{1}{4}\frac{b_{\wh z}}{\rho^2}
  -\frac{e^\phi}{8}\frac{c}{\rho^2}\right)\Gamma_{\wh r}\epsilon=0,\\
\Gamma_{\wh r}\delta\psi_{\wh r}
&=&\left(\frac{1}{2}\frac{h_{\wh z}}{h}
  -\frac{1}{4}\frac{b_{\wh r}}{\rho^2}
        +\frac{e^\phi}{8}m\right)\epsilon
  +\left(\frac{s_{\wh r}}{s}-\frac{e^\phi}{8}\frac{c}{\rho^2}\right)
                                                   \Gamma_{\wh r}\epsilon
        =0,\\
\Gamma_{\wh\theta^1}\delta\psi_{\wh\theta^1}
&=&\left(\frac{1}{2}\frac{\rho_{\wh z}}{\rho}
-\frac{1}{4}\frac{b_{\wh r}}{\rho^2}
+\frac{e^\phi}{8}m\right)\epsilon
\nonumber\\&&
+\left(-\frac{1}{2\rho}
       +\frac{1}{2}\frac{\rho_{\wh r}}{\rho}
       +\frac{1}{4}\frac{b_{\wh z}}{\rho^2}
       +\frac{e^\phi}{8}\frac{c}{\rho^2}\right)\Gamma^{\wh r}\epsilon=0,\\
\delta\lambda
&=&\left(
   \frac{s_{\wh z}}{s}
  +3\frac{f_{\wh z}}{f}
  +\frac{1}{2}\frac{h_{\wh z}}{h}
  +\frac{1}{2}\frac{\rho_{\wh z}}{\rho}
  -\phi_{\wh z}
  -\frac{1}{4}\frac{b_{\wh r}}{\rho^2}
                  \right)\epsilon
\nonumber\\&&
  +\left(
   \frac{s_{\wh r}}{s}
  +3\frac{f_{\wh r}}{f}
  +\frac{1}{2}\frac{g_{\wh r}}{g}
  +\frac{\rho_{\wh r}}{\rho}
  -\phi_{\wh r}
  -\frac{1}{\rho}
  +\frac{1}{4}\frac{b_{\wh z}}{\rho^2}
                  \right)\Gamma^{\wh r}\epsilon=0.
\end{eqnarray}
For the solution to maintain $1/4$ supersymmetry,
the coefficients of $\epsilon$ and $\Gamma_{\wh r}\epsilon$
in these equations should vanish individually.
This gives the following ten equations:
\begin{eqnarray}
&&\frac{f_z}{f}=-\frac{1}{4}mge^\phi,\hspace{3.6em}
\frac{f_r}{f}=\frac{1}{4}\frac{che^\phi}{\rho^2},
\label{feq}\\
&&\frac{s_z}{s}=-\frac{1}{8}mge^\phi,\hspace{3.8em}
\frac{s_r}{s}=\frac{1}{8}\frac{che^\phi}{\rho^2},
\label{seq}\\
&&\frac{\rho_z}{\rho}=\frac{1}{2}\frac{b_rg}{h\rho^2}-\frac{1}{4}mge^\phi,\hspace{1em}
\frac{\rho_r}{\rho}=\frac{h}{\rho}-\frac{1}{2}\frac{b_zh}{g\rho^2}-\frac{1}{4}\frac{che^\phi}{\rho^2},
\label{rhoeq}
\\
&&\phi_z=\frac{1}{2}\frac{b_rg}{h\rho^2}-\frac{5}{4}mge^\phi,\hspace{1.2em}
\phi_r=-\frac{1}{2}\frac{b_zh}{g\rho^2}+\frac{3}{4}\frac{che^\phi}{\rho^2},\\
&&\frac{h_z}{h}=\frac{1}{2}\frac{b_rg}{h\rho^2}-\frac{1}{4}mge^\phi,\label{hequ}\\
&&\hspace{10.8em}\frac{g_r}{g}=-\frac{1}{2}\frac{b_zh}{g\rho^2}+\frac{1}{4}\frac{che^\phi}{\rho^2}.
\end{eqnarray}

To fix the ambiguity of reparameterization as $x^i\rightarrow (\mbox{const})x^i$,
we impose the boundary condition
\begin{equation}
f=g=h=s=1,\quad
e^\phi=g_{\rm str}\quad
\mbox{at $P$}
\label{boundarycond}
\end{equation}
on these functions,
where $P$ is a specific point in the spacetime, which we can choose arbitrarily.
Although $P$ is ordinarily taken to be infinity,
we do not do so here, because the solutions we obtain
are not asymptotically flat.

Comparing (\ref{feq}) and (\ref{seq}), we find that $f$ is proportional to $s^2$.
The constant factor is fixed by the condition (\ref{boundarycond}),
and we have $f=s^2$.
Next, we can fix the freedom of reparameterization of the coordinate $r$
by setting $\rho=rh$.
This is always possible on a specific line of fixed $z$ in the $z$-$r$ plane,
and if we use such coordinates at some $z$,
the relation automatically holds for all $z$, because of the equation
\begin{equation}
\partial_z(\rho-rh)
=\left(\frac{1}{2}\frac{b_rg}{h\rho^2}-\frac{1}{4}mge^\phi\right)(\rho-rh),
\end{equation}
which we can obtain from (\ref{rhoeq}) and (\ref{hequ}).
Through this coordinate fixing,
$h_r$ is obtained from (\ref{rhoeq}) as
\begin{equation}
h_r=-\frac{1}{2}\frac{b_z}{gr^2}-\frac{1}{4}\frac{ce^\phi}{r^2}.
\end{equation}
Furthermore, we have the relation
\begin{equation}
\phi_z-8\frac{s_z}{s}-\frac{h_z}{h}=
\phi_r-8\frac{s_r}{s}-\frac{h_r}{h}=0.
\end{equation}
Thus, we can set $e^\phi=g_{\rm str}hs^8$.
The $r$-derivatives of the functions $h$, $g$ and $s$ satisfy
the relation
\begin{equation}
\frac{g_r}{g}-4\frac{s_r}{s}-\frac{h_r}{h}=0.
\end{equation}
This implies that $g/(hs^4)$ depends only on the coordinate $z$.
By a reparameterization of the coordinate $z$,
we can set $g=hs^4$.
Now, $f$, $\rho$, $\phi$ and $g$ are represented by two functions, $h$ and $s$.
Finally, we define the functions $S$ and $K$ by
\begin{equation}
K=h^2s^4,\quad
S=s^{-8}.
\end{equation}
Then we obtain the following differential equations for $K$ and $S$:
\begin{eqnarray}
&&S_z=mg_{\rm str}K,\quad
  S_r=-\frac{g_{\rm str}c}{r^2},
\label{eq50}
\\
&&(KS)_z=\frac{b_r}{r^2},\quad
  K_r=-\frac{b_z}{r^2}.
\label{eq51}
\end{eqnarray}
The relations (\ref{eq50}) are the same as (\ref{rr2}) and (\ref{KSeq}),
and (\ref{eq51}) reproduces the NS-NS $3$-form field strength (\ref{NSNS3}).
Using (\ref{eq50}), we can eliminate the function $K$ from (\ref{eq51}),
and we obtain
\begin{eqnarray}
-\frac{g_{\rm str}}{r^2}(c_z-mb_z)&=&0,\label{bianchi2}\\
-\frac{g_{\rm str}}{r^2}(c_r-mb_r)
&=&\frac{1}{r^2}\frac{\partial}{\partial r}r^2\frac{\partial}{\partial r}S
+\frac{1}{2}\frac{\partial^2}{\partial z^2}S^2.\label{bianchi1}
\end{eqnarray}
Because of the Bianchi identity $d(G_2-mB_2)=ddC_1=0$,
the left-hand sides of these equations vanish.
Therefore, Eqs. (\ref{bianchi2}) always holds, while
Eq. (\ref{bianchi1}) gives (\ref{Sequation}).

\section{D8-branes as domain walls}\label{domainwall.sec}
In this section, we construct D8-branes as domain wall solutions by
connecting two solutions with different values of the mass parameter.
We distinguish quantities in the two regions
by the labels `$+$' and `$-$',
indicating quantities on the $+z$ and $-z$ sides of the wall.
For example, mass parameters in the two regions are denoted by
$m^{(+)}$ and $m^{(-)}$.
Because the mass parameter jumps by one on each D8-brane,
a domain wall made of two solutions with $m^{(+)}$ and $m^{(-)}$ is
identified with $N=-(m^{(+)}-m^{(-)})$ coinciding D8-branes.
By looking at the $z$ dependence of the gravitational
potential $f$ in (\ref{feq}),
we find that $N>0$ ($N<0$) corresponds to positive (negative) tension branes.

From the equations of motion,
we obtain several boundary conditions on a domain wall.\cite{cpr}
From the equation of motion of the background metric,
we obtain the condition that
the induced metrics $\wt g_{mn}^{(\pm)}$
on the boundary made of the metrics of
two regions should coincide up to reparameterization.
Let us parameterize the domain wall world volume
by the coordinates $(x^i,r,\theta^m)$.
The shape of the domain wall is specified by a function $z=z(r)$.
There are two sets of worldvolume coordinates
defined by the coordinates on both sides of the domain wall.
The coordinates $x^i$ and $\theta^m$ are common to the two regions, and
the following ($+$) and ($-$) components of the induced metrics should coincide:
\begin{eqnarray}
\wt g^{(\pm)}_{ij}&=&S^{(\pm)-1/2}\eta_{ij},\\
\wt g^{(\pm)}_{\theta^m\theta^n}
                  &=&K^{(\pm)}S^{(\pm)1/2}r^{(\pm)2}\gamma_{\theta^m\theta^n}.
\end{eqnarray}
Precisely speaking, $\wt g^{(+)}_{ij}$ and $\wt g^{(-)}_{ij}$
can differ by a constant factor.
Here we take the point $P$ on the boundary,
and then the constant becomes $1$.
According to the equation of motion of
the dilaton field, it should also be continuous at the boundary,
and two values
\begin{equation}
e^{2\phi^{(\pm)}}=g_{\rm str}^2K^{(\pm)}S^{(\pm)-3/2}
\end{equation}
should coincide.
From the continuity of $\wt g_{ij}$ and $\phi$, we find that $K$ and $S$ are also continuous.
The condition for $\wt g_{\theta^m\theta^n}$ implies
that the coordinates $r^{(+)}$ and $r^{(-)}$
take the same values on the boundary.
Therefore, the $\wt g_{rr}$ components of the induced metrics,
$\wt g_{rr}^{(+)}$ and $\wt g_{rr}^{(-)}$,
\begin{eqnarray}
\wt g^{(\pm)}_{rr}=KS^{1/2}+KS^{-1/2}\left(\frac{dz^{(\pm)}}{dr}\right)^2
=KS^{1/2}\frac{1}{\cos^2\varphi^{(\pm)}},
\end{eqnarray}
should coincide.
Here $\varphi^{(\pm)}$ is the gradient angle of the boundary in the $z$-$r$ plane
defined by
\begin{equation}
\tan\varphi^{(\pm)}=\frac{1}{S^{1/2}}\frac{dz^{(\pm)}}{dr},\quad
|\varphi^{(\pm)}|<\frac{\pi}{2}.
\label{varphi}
\end{equation}
The coincidence of these quantities requires
the coordinate $z^{(\pm)}$ to also be continuous at the boundary.

From these facts,
the two sets of coordinates
$(x^{i(\pm)},z^{(\pm)},r^{(\pm)},\theta^{m(\pm)})$
on the domain wall coincide, and
the shape of a domain wall separating two regions is determined by
\begin{equation}
K^{(+)}(z,r)=K^{(-)}(z,r),\quad
S^{(+)}(z,r)=S^{(-)}(z,r).
\label{KKSS}
\end{equation}
In general, these two equations give different curves in the $z$-$r$ plane.
This implies that two such solutions cannot be connected by a domain wall.
To obtain a domain wall solution,
we need two solutions satisfying the two equations in (\ref{KKSS}) on
a single curve in the $z$-$r$ plane.

Unlike the metric and the dilaton field,
the field strength $G_2$ is not continuous at the boundary.
Its boundary condition is obtained from
the equation of motion of the R-R $7$-form field $C_7$.
The action of $C_7$ is
\begin{equation}
\frac{S}{2\pi}=\int\left(\frac{1}{2}G_8\wedge *G_8+N\delta_1\wedge{\cal F}_2\wedge C_7\right),
\end{equation}
where ${\cal F}_2=F_2+B_2$ and $\delta_1$ is the $1$-form $\delta$-function
representing the worldvolume of $N$ coinciding D8-branes.
We have assumed that the $\U(N)$ field strength on the branes is
proportional to the $N\times N$ unit matrix.
The equation of motion for $C_7$ is
\begin{equation}
dG_2=-N\delta_1\wedge{\cal F}_2,
\label{dgdf}
\end{equation}
where $G_2=*G_8$ is the field strength of the $1$-form R-R field.
Integrating (\ref{dgdf}) in an infinitesimal interval containing
the D8-branes, we obtain the boundary condition
\begin{equation}
-\frac{1}{N}(\wt G_2^{(+)}-\wt G_2^{(-)})
               ={\cal F}_2,
\label{bbc}
\end{equation}
where the tildes indicate the pullback onto the D8-branes.
Let us check the supersymmetry of this domain wall solution.
The fact that $dz^{(+)}/dr$ and $dz^{(-)}/dr$, or equivalently,
the gradient angles $\varphi^{(+)}$ and $\varphi^{(-)}$ of the boundary,
defined in the two regions take the same value
implies that the local Lorentz frame can be connected smoothly
on the boundary.
Therefore, the parameter $\epsilon$ defined by the condition (\ref{susycond})
is defined continuously, and this domain wall solution,
excluding the boundary itself, is a BPS configuration.

For the boundary itself, we should check the BPS condition separately,
using analysis similar to that for BIon configurations
in Ref.\citen{CM}.
In our system, only one component, ${\cal F}_{\theta^1\theta^2}$, of the
$\U(1)$ gauge field can be non-zero,
because of the symmetry which we have assumed.
The BPS condition for the D8-brane is
\begin{equation}
(\cos\varphi\Gamma_{\wh z}+\sin\varphi\Gamma_{\wh r})
(\cos\psi+\sin\psi\Gamma_{\wh \theta^1\wh\theta^2}\Gamma_{\wh{11}})\epsilon=\epsilon,
\end{equation}
where we have introduced the angle $\psi$ by
\begin{equation}
\tan\psi={\cal F}_{\wh\theta^1\wh\theta^2},\quad
|\psi|<\frac{\pi}{2}.
\label{magnetic}
\end{equation}
For any $\epsilon$ satisfying the condition (\ref{susycond})
also to satisfy this equation,
the relation
\begin{equation}
\varphi-\psi=0
\label{phipsi}
\end{equation}
must hold.
We can show that this condition is satisfied
on the boundary of our domain wall solutions as follows.
Differentiating the second equation in (\ref{KKSS}), we obtain
\begin{equation}
(S^{(+)}_z-S^{(-)}_z)dz+(S^{(+)}_r-S^{(-)}_r)dr=0.
\end{equation}
With the relations (\ref{rr2}) and (\ref{KSeq}),
this can be rewritten as
\begin{equation}
\frac{1}{S^{1/2}}\frac{dz}{dr}-
\frac{G_{\wh\theta^1\wh\theta^2}^{(+)}-G_{\wh\theta^1\wh\theta^2}^{(-)}}{m^{(+)}-m^{(-)}}=0.
\end{equation}
Using Eq. (\ref{varphi})
and the boundary condition (\ref{bbc}), this is reduced to
\begin{equation}
\tan\varphi-\tan\psi=0.
\end{equation}
Taking account of the fact that $|\varphi|,|\psi|<\pi/2$,
this is equivalent to the BPS condition (\ref{phipsi}).

Before ending this section, we would like to mention orientifold planes.
Solutions including (positive tension) D8-branes always have singularities with $S=0$ located
at a finite distance in the $z$-direction.
On the singularities, the dilaton field diverges,
and we cannot treat them perturbatively.
To avoid them, we can use orientifold $8$-planes.
In the context of supergravity,
we can regard O8-planes as 8-branes with negative tension
positioned at fixed points
under the mapping $z\rightarrow-z$
whose ${\bf Z}_2$ symmetries
are as listed in Table \ref{o8parity.tbl}.
\begin{table}[htb]
\caption{Intrinsic parities of the bosonic fields under the ${\bf Z}_2$ action associated with
orientifold 8-planes.}
\label{o8parity.tbl}
\begin{center}
\begin{tabular}{ccccccccc}
\hline
\hline
$g_{mn}$ & $B_2$ & $\phi$ & $G_0$ & $G_2$ & $G_4$ & $G_6$ & $G_8$ & $G_{10}$ \\
\hline
$+$ & $-$ & $+$ & $-$ & $+$ & $-$ & $+$ & $-$ & $+$ \\
\hline
\end{tabular}
\end{center}
\end{table}
Due to this ${\bf Z}_2$ symmetry, the field strength $G_2$ on the two sides of an orientifold plane
coincide, and the condition (\ref{bbc}) is satisfied if ${\cal F}_2=0$.
This is consistent with the fact that there is no gauge field on orientifold planes.

\section{Examples}\label{examples.sec}
We now give several concrete solutions below as examples.

\subsection{Massless solutions}
Massless solutions are specified by two independent functions $K$ and $S$
associated with NS5-branes and D6-branes, respectively.
These functions are solutions of differential equations
(\ref{Sequation}) and (\ref{Kequation}).
The function $S$ is determined easily as (\ref{masslessS}).
There are several cases in which we can also obtain the function $K$ analytically.
\subsubsection{The case of pure D6-branes}
A configuration consisting of only coinciding D6-branes is given by
\begin{equation}
K=1,\quad
S=1+\frac{Q_6g_{\rm str}}{4\pi r},
\end{equation}
where $Q_6$ is the D6-brane charge.
\subsubsection{The case of pure NS5-branes}
If $Q_6=0$ and $S=1$, the differential equation (\ref{Kequation}) reduces
to the Laplace equation on the flat ${\bf R}^4$, which can be easily solved.
For example, $Q_5$ NS5-branes at $z=r=0$ are given by
\begin{equation}
K=1+\frac{Q_5g_{\rm str}}{2\pi^2(z^2+r^2)},\quad
S=1.
\end{equation}
\subsubsection{The case of D6-branes with smeared NS5-branes}
If the function $K$ does not depend on $z$,
(\ref{Kequation}) can be solved easily by
\begin{equation}
K=1+\frac{\rho_5}{4\pi r},\quad
S=1+\frac{Q_6g_{\rm str}}{4\pi r}.
\label{NS5D6sol}
\end{equation}
This represents $Q_6$ D6-branes with smeared NS5-branes on them.
The constant $\rho_5$ is the NS5-brane charge density.

\subsection{Massive solutions}
If $m\neq0$, the function $K$ is uniquely determined by the function $S$
through Eq. (\ref{KSeq}).
This is because the NS5-brane charge density is not independent of
the D6-brane charge density.
Therefore, the solution is specified by only the function $S$.
The function $S$ is a solution
of the differential equation (\ref{Sequation}).
Because this equation is non-linear,
it is difficult in general to solve analytically.
However, we can obtain general solutions order by order in the following way.
Let us expand the function $S$ with respect to $r$ as
\begin{equation}
S(z,r)=\sum_{n=-1}^{\infty}a_n(z)r^n.
\end{equation}
By substituting this into (\ref{Sequation}), we obtain
\begin{equation}
n(n+1)a_{n}=-\frac{1}{2}\frac{\partial^2}{\partial z^2}
           \sum_{k=0}^{n}a_{k-1}a_{n-k-1}.
\label{eq15}
\end{equation}
For $n=0$, this equation reduces to $\partial_z^2a_{-1}^2=0$, and
no constraint is imposed on the function $a_0$.
All $a_n$ with $n\geq1$ can be found in terms of $a_{-1}$ and $a_0$
using (\ref{eq15}) order by order.
In general, this procedure does not end at finite order.
However, there are (at least) two cases in which
only a finite number of coefficients $a_n$ do not vanish.

\subsubsection{The case of massive D6-branes}
A massive D6-brane solution is given by
\begin{equation}
S=(1+2mg_{\rm str}z)^{1/2}\left(1+\frac{Q_6g_{\rm str}}{4\pi r}\right).
\label{D8sol}
\end{equation}
Even in the massless limit, $m\rightarrow0$, this does not reproduce the pure D6-brane solution.
Instead, it converges to the D6-NS5 solution (\ref{NS5D6sol})
with $\rho_5=g_{\rm str}Q_6$.
If $Q_6=0$, this solution reduces to a solution given in
Refs.\citen{PolWit}and \citen{7and8}.
\subsubsection{The case of massive D6-branes on a $B\neq0$ background}
Another kind of massive D6-brane solutions has been obtained by
Janssen et al.\cite{janssen}
It is the following:
\begin{equation}
S=1+mg_{\rm str}z-\frac{1}{6}m^2g_{\rm str}^2r^2+\frac{Q_6g_{\rm str}}{4\pi r}.
\label{Janssensol}
\end{equation}
The constant $Q_6$ represents the D6-brane charge.
Even in case $Q_6=0$,
this solution does not coincide with (\ref{D8sol}).
In particular, the $B$-field of this solution becomes larger as $r$ grows.
Therefore, this is identified with a solution for which the NS5-brane charge
is put at infinity.

\subsection{D8-branes}
Following to the prescription given in section \ref{domainwall.sec},
we can construct several D8-brane solutions.

\subsubsection{The D6-D8 solution}
Let us use two solutions in the form (\ref{D8sol}) with different $m$.
The two equations (\ref{KKSS}) are satisfied simultaneously
only if $Q_6^{(+)}=Q_6^{(-)}$.
In this case, the domain wall is the plane $z=0$.

\subsubsection{The curved D8-branes}
It is also possible to use the solution (\ref{Janssensol})
to construct domain walls.
Because the function $K$ is always $1$ in this solution,
we can construct domain wall solutions for any
values of $Q_6^{(\pm)}$.
Here, let us discuss a solution with vanishing D6-brane charge, $Q_6^{(\pm)}=0$.
The equation $S^{(+)}=S^{(-)}$ is solved by
\begin{equation}
z=\frac{g_{\rm str}}{6}(m^{(+)}+m^{(-)})r^2.
\label{curvedD8}
\end{equation}
This is a curved D8-brane solution.
Curved domain wall solutions have recently been discussed
in the literature\cite{curved,ricchiflat,more}.
What we have presented above, however, is different.
In fact, the D8-brane world volume (\ref{curvedD8})
is curved parabolically, and is not Ricchi flat.
\subsubsection{The BIon solution}
As mentioned above, we can use the two solutions (\ref{Janssensol}) with
arbitrary $Q_6^{(\pm)}$ to construct other D8-brane configurations.
The curve defined by $S^{(+)}=S^{(-)}$ is
\begin{equation}
z=\frac{g_{\rm str}}{6}(m^{(+)}+m^{(-)})r^2
            -\frac{Q_6^{(+)}-Q_6^{(-)}}{m^{(+)}-m^{(-)}}\frac{1}{4\pi r}.
\end{equation}
In particular, if one of the charges $Q_6^{(\pm)}$ is zero
and the other is negative,
this is a supergravity realization of the BIon configuration\cite{CM}.

\section{D8-D0-F1 solution}\label{018.sec}
Systems consisting of
D-particles, fundamental strings along the $z$ direction
and D8-branes perpendicular to the $z$ direction,
are also 1/4 BPS.
The unbroken supersymmetries are specified by the conditions
\begin{equation}
\Gamma^z\epsilon=\epsilon,\quad
\Gamma^{0z}\Gamma_{11}\epsilon=\epsilon.
\end{equation}
\begin{table}[htb]
\caption{$1/4$ BPS configuration consisting of D8, D0 and F1-branes}
\label{018.tbl}
\begin{center}
\begin{tabular}{ccccccccccc}
\hline
\hline
           & $x^0$ & $z$ & $r$ & $\theta^1$ & $\theta^2$ & $\theta^3$ & $\theta^4$ & $\theta^5$ & $\theta^6$ & $\theta^7$ \\
\hline
D8 & $\circ$ && $\circ$ & $\circ$ & $\circ$ & $\circ$ & $\circ$ & $\circ$ & $\circ$ &$\circ$ \\
D0 & $\circ$ \\
F1 & $\circ$ & $\circ$ \\
\hline
\end{tabular}
\end{center}
\end{table}

In Ref.\citen{081}, employing a method similar to ours,
the authors represent the solution with two functions.
In the massive case, however,
these two functions are not independent, and
we can represent the solution using only one function, $S$.
This solution is
\begin{equation}
ds^2=-\frac{1}{SK^{1/2}}(dx^0)^2+\frac{K^{1/2}}{S}dz^2
     +K^{1/2}(dr^2+r^2d\Omega_7^2),
\label{018metric}
\end{equation}
\begin{equation}
e^\phi=g_{\rm str}\frac{K^{3/4}}{S^{1/2}},\quad
C_0=\frac{1}{g_{\rm str}K},\quad
B_{0z}=\frac{1}{S}.
\label{018others}
\end{equation}
Here $K$ and $S$ are functions of $z$ and $r$.
If $m\neq0$, the function $K$ is determined by the function $S$ through
the relation
\begin{eqnarray}
mg_{\rm str}K=\frac{\partial S}{\partial z}.
\label{018KSeq}
\end{eqnarray}
From the equation of motion of $B_{0z}$, we obtain
\begin{equation}
\Delta_8S+\frac{1}{2}\frac{\partial^2}{\partial z^2}S^2=0,
\label{018Seq}
\end{equation}
where $\Delta_8$ is the Laplacian in the $(r,\theta^m)$ plane:
\begin{equation}
\Delta_8=\frac{1}{r^7}\frac{\partial}{\partial r}r^7\frac{\partial}{\partial r}.
\label{lap8}
\end{equation}
Note that the corresponding equation in the D6-NS5 system is obtained from the Bianchi identities
for $B_2$ and $C_1$.
This difference results from the fact that $B_2$ and $C_1$ are electrically coupled
to the F1 and D0-branes, while they are magnetically coupled to the NS5 and D6-branes.

In the massless case, the function $K$ is determined by
\begin{equation}
\Delta_8K+S\frac{\partial^2}{\partial z^2}K=0,
\label{018Keq}
\end{equation}
instead of (\ref{018KSeq}).

By solving (\ref{018Seq}) [and (\ref{018Keq}) in the massless case],
we can obtain many kinds of solutions appearing in Ref.\citen{081}.
Domain wall solutions are constructed in the same way as in
the D8-D6-NS5 systems.
The connecting conditions are
\begin{equation}
K^{(+)}(z,r)=K^{(-)}(z,r),\quad
S^{(+)}(z,r)=S^{(-)}(z,r).
\end{equation}
These two conditions must hold simultaneously on a domain wall.
The BPS  condition on D8-branes is shown to hold as follows.
Let us introduce two angles
\begin{equation}
\tan\varphi=\frac{1}{S^{1/2}}\frac{dz}{dr},\quad
\sin\psi={\cal F}_{\wh0\wh r},\quad
|\varphi|,|\psi|<\frac{\pi}{2}.
\end{equation}
[Note that, in the contrast to the magnetic component (\ref{magnetic}),
the electric component ${\cal F}_{\wh0\wh r}$ never exceeds
the critical value $1$.\cite{OM1,OM2}]
The BPS condition is
\begin{equation}
\varphi=\psi.
\label{susyond8in018}
\end{equation}
From the boundary condition (\ref{bbc}), we have
\begin{equation}
{\cal F}_{\wh0\wh r}=\frac{\wt G_{\wh0\wh r}-\wt G_{\wh0\wh r}}{m^{(+)}-m^{(-)}}
             =\wt B_{\wh0\wh r}.
\end{equation}
From the solution represented by (\ref{018metric}) and (\ref{018others}),
the pull-back of the NS-NS $2$-form field on the domain wall is
found to be
\begin{equation}
\wt B_{\wh 0\wh r}=\sin\varphi.
\end{equation}
From these equations, we find that Eq. (\ref{susyond8in018}) holds.

\section{D8-D2 and D8-D4 solutions}\label{2848.sec}
There are two more kinds of 1/4 BPS solutions,
D8-D2 and D8-D4 solutions.
The D8-D2 solution and the delocalized D8-D4
solution are constructed in Ref.\citen{bbj}.
The localized version of the D8-D4 solution is obtained in \citen{youm}
and is used in Ref.\citen{bo} for analysis
of the fixed points of five-dimensional
supersymmetric gauge theories.
The D8-D4 solution is also obtained in Ref.\citen{F4}
in a different form.
The orientation of each brane is given in Tables \ref{28.tbl} and \ref{48.tbl}.
\begin{table}[htb]
\caption{1/4 BPS configuration consisting of D8 and D2-branes.}
\label{28.tbl}
\begin{center}
\begin{tabular}{ccccccccccc}
\hline
\hline
& $x^0$ & $x^1$ & $z$ & $r$ & $\theta^1$ & $\theta^2$
 & $\theta^3$ & $\theta^4$ & $\theta^5$ & $\theta^6$ \\
\hline
D8-brane & $\circ$ & $\circ$ && $\circ$ & $\circ$ & $\circ$ & $\circ$ & $\circ$ & $\circ$ & $\circ$ \\
D2-brane & $\circ$ & $\circ$ & $\circ$ \\
\hline
\end{tabular}
\end{center}
\end{table}
\begin{table}[htb]
\caption{1/4 BPS configuration consisting of D8 and D4-branes.}
\label{48.tbl}
\begin{center}
\begin{tabular}{ccccccccccc}
\hline
\hline
& $x^0$ & $x^1$ & $x^2$ & $x^3$ & $x^4$ & $z$ & $r$ & $\theta^1$ & $\theta^2$
 & $\theta^3$ \\
\hline
D8-brane & $\circ$ & $\circ$ & $\circ$ & $\circ$ & $\circ$ && $\circ$ & $\circ$ & $\circ$ & $\circ$ \\
D4-brane & $\circ$ & $\circ$ & $\circ$ & $\circ$ & $\circ$ \\
\hline
\end{tabular}
\end{center}
\end{table}
These solutions can be reproduced
in a manner similar to that used in the previous section.
We find the D8-D2 solution to be
\begin{equation}
ds^2=H_8^{-1/2}H_2^{-1/2}\eta_{ij}dx^idx^j+H_8^{1/2}H_2^{-1/2}dz^2+H_8^{-1/2}H_2^{1/2}(dr^2+r^2d\Omega_6^2),
\end{equation}
\begin{equation}
e^\phi=g_{\rm str}H_8^{-5/4}H_2^{1/4},\quad
C_{01z}=\frac{H_8}{g_{\rm str}H_2},
\end{equation}
and the D8-D4 solution to be
\begin{equation}
ds^2=H_8^{-1/2}H_4^{-1/2}\eta_{ij}dx^idx^j+H_8^{1/2}H_4^{1/2}dz^2+H_8^{-1/2}H_4^{1/2}(dr^2+r^2d\Omega_3^2),
\end{equation}
\begin{equation}
e^\phi=g_{\rm str}H_8^{-5/4}H_4^{-1/4},\quad
C_{01234}=\frac{1}{g_{\rm str}H_4},
\end{equation}
where the $5$-form R-R potential $C_5$ is defined by $dC_5=-*G_4$.
Both of these solutions can be written in terms of two harmonic functions,
$H_8$ and $H_{2/4}$.
The function $H_8$ is associated with D8-branes and depends only
on the coordinate $z$.
It satisfies
\begin{equation}
\frac{d}{dz}H_8(z)=mg_{\rm str}.
\end{equation}
$H_2$ and $H_4$ are harmonic functions associated with
the D2-branes and D4-branes, respectively.
They are solutions of the Laplace equations
\begin{equation}
\Delta_7H_2(r)=0,\quad
\Delta_4H_4(z,r)+\frac{1}{H_8(z)}\frac{\partial^2}{\partial z^2}H_4(z,r)=0,
\end{equation}
where $\Delta_7$ and $\Delta_4$ are the Laplacians in the $(r,\theta^m)$ planes.
Because we assume rotational symmetry,
these include only $r$-derivative terms:
\begin{equation}
\Delta_7=\frac{1}{r^6}\frac{\partial}{\partial r}r^6\frac{\partial}{\partial r},\quad
\Delta_4=\frac{1}{r^3}\frac{\partial}{\partial r}r^3\frac{\partial}{\partial r}.
\label{lap74}
\end{equation}
Because of D2-brane charge conservation,
$H_2$ does not depend on $r$, while $H_4$ is a function of both
$z$ and $r$.

D8-branes are again introduced as domain walls
separating regions with different values of $m$.
The matching conditions on the walls are
\begin{equation}
H_8^{(+)}=H_8^{(-)},\quad
H_{2/4}^{(+)}=H_{2/4}^{(-)}.
\label{h824cond}
\end{equation}
Because $H_8$ is a function of $z$,
the solution for the first condition in (\ref{h824cond})
is $z={\rm(constant)}$.
The condition for $H_2$ implies that the D2-brane charge does not
vary on D8-branes and that D2-branes cannot end on D8-branes.

Unlike the D8-D0-F1 and D8-D6-NS5 cases,
the differential equations for $H_2$ and $H_4$ are linear.
This may be related to the fact that brane creation through the Hanany-Witten
effect does not occur in D2-D8 or D4-D8 systems.

\section{Discussions}
In this paper, with the help of supersymmetry conditions,
we construct
four kinds of $1/4$ BPS solutions in massive IIA supergravity:
D8-D0-F1, D8-D2, D8-D4 and D8-D6-NS5 solutions.
In obtaining these solutions,
we assumed rotational symmetries in the $(r,\theta^m)$ planes.
Considering the solutions we have obtained,
we can easily guess that more general solutions
without rotational symmetry
may be obtained by replacing the Laplacians (\ref{laplacian}), (\ref{lap8})
and (\ref{lap74}) by the full Laplacians.

Among the four solutions,
the D8-D2 and D8-D4 solutions are represented by two harmonic functions
associated with the two kinds of branes in each solution.
However, the other two solutions, consisting of three kinds of branes,
have different structure.
In massive regions, each of them is represented by only one function,
which is a solution of a non-linear differential equation.
This difference is related to the correlation of brane charges
in massive IIA supergravity.

The supergravity solutions we have discussed may be used
for analysis of dynamics of gauge theories.
Here, let us discuss the application of the D8-D6-NS5 solutions.
It would be possible to study three- and two-dimensional gauge theories
using the D2-brane probes in our solutions.
We can position D2-branes along $x^0$, $x^1$ and $z$ directions
in the D8-D6-NS5 solution as probes.
The energy density of this D2-brane is
\begin{equation}
\frac{2\pi}{e^\phi}\sqrt{-g_{00}g_{11}g_{zz}}=\frac{2\pi}{g_{\rm str}}.
\end{equation}
Because this is constant and independent of the coordinates,
the D2-brane is stable at any point in the $(x^2,x^3,x^4,x^5)$ plane.

Many simple solutions included in our solution have been
previously used for analysis of gauge theories.
If we consider a massless background with only D6-branes,
we obtain ${\cal N}=4$ $\U(1)$ gauge theory on a D2-brane.
If we add an orientifold 6-plane,
the gauge group becomes $\SU(2)$.
Nonperturbative properties of these theories
can be understood by lifting the configurations
up to M-theory configurations.\cite{s3dim,sw3dim}

In order to study two-dimensional gauge theories, let us compactify
the $z$-direction on ${\bf S}^1$ with period $L$.
If we add NS5-branes as the background,
the probe D2-brane splits into segments,
ending on the NS5-branes,
and we obtain Hanany-Witten type configurations.
This system possesses ${\cal N}=(4,4)$ supersymmetry and
is investigated in Ref.\citen{HH}.
When we introduce D6 and/or D8-branes,
supersymmetry is broken to ${\cal N}=(4,0)$.
If we introduce only D6-branes and no D8-brane,
we obtain the gauge theories discussed in Ref.\cite{40}.
The moduli space of the Coulomb branch of this theory
is ${\bf R}^4\times{\bf S}^3/{\bf Z}_{Q_5}\times{\bf R}$\cite{40}, and
it is obtained by a D2-brane probe as follows.
In the static gauge $X^\mu=\sigma^\mu$,
the Born-Inferd action of the D2-brane in the D6-NS5 solution (\ref{NS5D6sol})
is
\begin{eqnarray}
\frac{S}{2\pi}&=&\frac{L}{g_{\rm str}}
     \left(
       \frac{1}{4}S{\cal F}_{\mu\nu}^2
       +\frac{1}{2}(\partial_\mu x^i)^2
\right.
\nonumber\\&&
\left.
       +\frac{1}{2}\frac{S}{K}(\partial_\mu A_z+\rho_5{\cal A}_{\theta^m}\partial_\mu\theta^m)^2
       +\frac{1}{2}KS(\partial_\mu y^a)^2\right),
\end{eqnarray}
where ${\cal A}_{\theta^m}$ is a gauge potential on ${\bf S}^2$ surrounding a
monopole carrying unit charge.
In the small $L$ limit,
we obtain the  metric on the moduli space as
\begin{equation}
ds^2=\frac{Q_5L}{g_{\rm str}}(dx^i)^2
       +\frac{Q_5}{\pi g_{\rm str}}S
         (d\rho^2+\rho^2d\Omega_{3/Q_5}^2),
\end{equation}
where
\begin{equation}
d\Omega_{3/Q_5}^2
=(2\pi)^2\left(\frac{da}{Q_5}+{\cal A}_{\theta^m}\partial_\mu\theta^m\right)^2
       +\frac{1}{4}d\Omega_2^2
\end{equation}
is the metric of the lens space ${\bf S}^3/{\bf Z}_{Q_5}$
and we have introduced the scalar fields $a=LA_z$ and $\rho=r^{1/2}$.
This is the moduli space of the Coulomb branch of the $(4,0)$ theory\cite{40}.
If $Q_5=1$, ${\cal N}=(4,4)$ supersymmetry is restored
in the limit $L\rightarrow0$
and the results of the $A_n$-type theory in Ref.\citen{DiaSei} are reproduced.

If we introduce D8-branes and O8-planes,
we obtain different ${\cal N}=(4,0)$ gauge theories.
An interesting question is what in two-dimensional gauge theory
corresponds to the variety of solutions we have obtained.
In particular, the physical implication of the existence of the two solutions
(\ref{D8sol}) and (\ref{Janssensol})
(and the infinite number of solutions interpolating between them)
is mysterious.
This seems to correspond to some kind of deformation of two-dimensional gauge theory.
The solution (\ref{Janssensol}) includes a non-zero $H_3$,
which appears to be the mass deformation of ${\cal N}=4$ theory to ${\cal N}=1^\ast$
studied in Ref.\citen{PolStr}.
However, this is not the case,
because the D2-brane can be moved freely in the $(x^2,x^3,x^4,x^5)$ plane,
as mentioned above, and corresponding scalar modes do not become massive.
Furthermore, we can show that
the D2-brane is not puffed up into a spherical D4-brane\cite{myers} in our background.
In the solution (\ref{Janssensol}),
even if we introduce a cutoff at some $z$ with orientifold planes,
there is a singular surface with $S=0$ at $r\sim1/(m^2g_{\rm str}^2)$
unless $m=0$.
In the five-dimensional gauge theory on a D4-brane probe
discussed in Refs.\citen{5dim1} and \citen{5dim2},
the emergence of such singularities is related to the existence of
a non-trivial conformal fixed point in five-dimensional gauge theories.
The singularity in the solution (\ref{Janssensol}) may
be related to such phenomena.

\section*{Acknowledgements}
The author would like to thank K. Ohta and T. Yokono
for discussions.

\appendix
\section{Conventions and notations}\label{convention.sec}
In this paper, we use the convention in which $\alpha'=1/(2\pi)^2$.
This implies that the string tension is $2\pi$.
For the signature of the metric, gamma matrices and spinors,
we use the convention used by Romans\cite{massiveiia}.
We redefine the supergravity fields as follows.
(The superscript ``R'' indicates that the field in question is that used in
Ref.\citen{massiveiia}.)
The Lagrangian is rescaled by $L^{\rm R}=(1/8\pi)L$.
The Weyl rescaling of the metric and the redefinition of the dilaton field is
given by
\begin{equation}
g^{\rm R}_{mn}=\xi^2g_{mn},\quad
\phi^{\rm R}=-\frac{1}{2}\phi,
\end{equation}
with $\xi=\exp(-\phi/4)=\exp(\phi^{\rm R}/2)$.
The anti-symmetric tensor fields are rescaled as
\begin{eqnarray}
2F^{\rm R}_4&=&G_4=dC_3+\frac{m}{2}B_2\wedge B_2,\\
2G^{\rm R}_3&=&H_3=dB_2,\\
2mB^{\rm R}_2&=&G_2=dC_1+mB_2.
\end{eqnarray}
In our convention,
the unit of the flux quantization of these field strengths is $1$.
The relations among the fermionic fields are
\begin{equation}
\psi^{\rm R}_m=\frac{\xi^{1/2}}{2}\left(\psi_m-\frac{1}{8}\Gamma_m\Gamma^n\psi_n
                                                       +\frac{1}{8}\Gamma_m\lambda\right),\quad
\lambda^{\rm R}=\frac{\xi^{-1/2}}{4\sqrt2}\left(\lambda-\Gamma^m\psi_m\right).
\end{equation}
Also, the supersymmetry parameter is rescaled as
\begin{equation}
\epsilon^{\rm R}=\frac{\xi^{1/2}}{2}\epsilon.
\end{equation}
With these conventions, the Lagrangian is given as a sum of the following parts:
\begin{eqnarray}
(2\pi e)^{-1}L_b
&=&\frac{1}{e^{2\phi}}
   \left(-R-4(\partial\phi)^2+\frac{1}{2\cdot3!}H_3^2\right)
\nonumber\\&&
   +\frac{1}{2}m^2
   +\frac{1}{2\cdot2}G_2^2
   +\frac{1}{2\cdot4!}G_4^2,\\
(2\pi e)^{-1}L_{f}
&=&\frac{1}{e^{2\phi}}
         \left(\frac{1}{2}\psi^m\Gamma^n\nabla_n\psi_m
                -\psi^n\nabla_n\lambda
           -\frac{1}{2}\lambda\Gamma^n\nabla_n\lambda\right),\\
(2\pi e)^{-1}L_{fGf}
&=&\frac{1}{16e^\phi}
         \left(-\psi_m\Gamma^n\sla G\Gamma^m\psi_n
                         -\lambda\sla G\lambda\right),\\
(2\pi e)^{-1}L_{fHf}
&=&\frac{1}{e^{2\phi}}\left(
 -\frac{1}{16}\psi_m\Gamma^m\sla H_3\Gamma^n\Gamma_{11}\psi_n
 +\frac{1}{16}\psi_m\Gamma^n\sla H_3\Gamma^m\Gamma_{11}\psi_n
\right.\nonumber\\&&\left.
 -\frac{1}{8}\psi_m\sla H_3\Gamma^m\Gamma^n\Gamma_{11}\psi_n
 -\frac{1}{8}\psi_m\Gamma^m\sla H_3\Gamma_{11}\lambda
\right.\nonumber\\&&\left.
 -\frac{1}{8}\psi_m\sla H_3\Gamma^m\Gamma_{11}\lambda
 -\frac{1}{8}\lambda\sla H_3\Gamma_{11}\lambda
 \right),\\
(2\pi e)^{-1}L_{\rm top}
&=&-\frac{1}{2}G_4G_4B_2
             +\frac{m}{3}G_4B_2B_2B_2
             -\frac{m^2}{15}B_2B_2B_2B_2B_2
\nonumber\\
&=&-\frac{1}{2}G^0_4G^0_4B_2
             -\frac{m}{6}G^0_4B_2B_2B_2
             -\frac{m^2}{40}B_2B_2B_2B_2B_2,
\end{eqnarray}
where $\sla{H}_3$ and $\sla{G}$ are as follows:
\begin{equation}
\sla{H}_3=\frac{1}{3!}H_{mnp}\Gamma^{mnp},\quad
\sla{G}=m+\frac{1}{2}G_{mn}\Gamma^{mn}\Gamma_{11}
           +\frac{1}{4!}G_{mnpq}\Gamma^{mnpq}.
\end{equation}

\end{document}